\documentclass[preprint2]{aastex}

\newcommand{\msun}{{\rm M_\odot}}
\newcommand{\hoz}{\rm H_2 ~ 1\!-\!0 ~ S(1)}

\newcommand{\htos}{\rm 2\!-\!1 ~ S(1)}
\newcommand{\kms}{\rm km\;s^{-1}}

\newcommand{\intenwat}{\rm W\;m^{-2}\;arcsec^{-2}}

\newcommand{\intennht}{\rm Jy\;beam^{-1}\;km\;s^{-1}}
\newcommand{\sintenwat}{\rm W\;m^{-2}\;arcsec^{-2}\;km^{-1}\;s}

\newcommand{\sintennht}{\rm Jy\;beam^{-1}}

\slugcomment{Accepted for publication in ApJ Jan 20, 2008, v673
n1.}

\shorttitle{3-D H$_2$ Observations around Sgr A East}
\shortauthors{Lee et al.}

\begin{document}

\title{Three-Dimensional Observations of H$_2$ Emission \\
    around Sgr A East \\
  - I. Structure in the Central 10 Parsecs of Our Galaxy}

\author{
    Sungho Lee\altaffilmark{1},
    Soojong Pak\altaffilmark{2},
    Minho Choi\altaffilmark{1},
    Christopher J. Davis\altaffilmark{3},
    T. R. Geballe\altaffilmark{4}, \\
    Robeson M. Herrnstein\altaffilmark{5,6},
    Paul T. P. Ho\altaffilmark{5,7},
    Y. C. Minh\altaffilmark{1,7},
    and
    Sang-Gak Lee\altaffilmark{8}
}

\altaffiltext{1}{Korea Astronomy and Space Science Institute,
     61-1 Hwaam-dong, Yuseong-gu, Daejeon 305-348, South Korea}
\altaffiltext{2}{Department of Astronomy and Space Science,
    Kyung Hee University, Yongin-si, Gyeonggi-do 446-701, South Korea;
    Corresponding Author, soojong@khu.ac.kr}
\altaffiltext{3}{Joint Astronomy Centre, University Park,
     660 North A'ohoku Place, Hilo, HI 96720}
\altaffiltext{4}{Gemini Observatory,
    670 N. A'ohoku Place, Hilo, HI 96720}
\altaffiltext{5}{Harvard-Smithsonian Center for Astrophysics,
     60 Garden Street, Cambridge, MA 02138}
\altaffiltext{6}{Department of Astronomy, Columbia University,
     550 West 120th St. New York, NY 10027}
\altaffiltext{7}{Academia Sinica Institute of Astronomy and
    Astrophysics, P.O. Box 23-141, Taipei, 106 Taiwan}
\altaffiltext{8}{Department of Astronomy, Seoul National
University,
     Kwanak-gu, Seoul 151-742, South Korea}

\begin{abstract}
  We have obtained velocity-resolved spectra of the
$\hoz ~ (\lambda = 2.1218 \micron)$ emission line at $2\arcsec$
angular resolution (or $\sim 0.08$~pc spatial resolution) in four
regions within the central 10 pc of the Galaxy where the
supernova-like remnant Sgr A East is colliding with molecular
clouds. To investigate the kinematic, physical, and positional
relationships between the important gaseous components in the
center, we compared the H$_2$ data cube with previously published
NH$_3$ data. The projected interaction-boundary of Sgr A East is
determined to be an ellipse with its center offset $\sim1.5$~pc
from Sgr A* and dimensions of 10.8~pc $\times$ 7.6~pc. This H$_2$
boundary is larger than the synchrotron emission shell but
consistent with the dust ring which is believed to trace the
shock front of Sgr A East. Since Sgr A East is driving shocks into
its nearby molecular clouds, we can determine their positional
relationships using the shock directions as indicators. As a
result, we suggest a revised model for the three-dimensional
structure of the central 10~pc. The actual contact between Sgr A
East and all of the surrounding molecular material, including the
circum-nuclear disk and the southern streamer, makes the
hypothesis of infall into the nucleus and feeding of Sgr A* very
likely.
\end{abstract}


\section{INTRODUCTION}

In the central 10 pc of our Galaxy, the Sgr A region contains
several characteristic objects; a candidate for super-massive
black hole (Sgr A*) of about $4 \times 10^6~\msun$ (see
\citealt*{ghe03,sch03} and references therein), a surrounding
cluster of stars (the Central cluster), molecular and ionized gas
clouds (the circum-nuclear disk (CND) and Sgr A West), supernova
remnants (SNR G~359.92-0.09 and Sgr A East). They are surrounded
by molecular structures including two giant molecular clouds
(GMCs) M-0.02-0.07 and M-0.13-0.08 (also known as the `$50~\kms$
cloud' and the `$20~\kms$ cloud', respectively).  In addition to
the two GMCs, recent accurate radio observations have resolved
several dense and filamentary molecular features around the Sgr A
complex; the `molecular ridge', the `southern streamer', the
`northern ridge', and the `western streamer' (see
Figure~\ref{3D_model} of this paper and Figures~3, 9, and 10 of
\citealt*{mcg01} and Figure~14 of \citealt*{her05}). These
molecular features are believed to play important roles in feeding
the central massive black hole \citep{ho91,coi99,coi00,mcg01}. The
interaction between these various components is responsible for
many of the phenomena occurring in this complicated and unique
portion of the Galaxy. Developing a comprehensive picture of the
primary interactions between the components at the Galactic center
will also improve our understanding of the nature of galactic
nuclei in general.

As the complicated morphology of the central 10 pc is being
unveiled thanks to the dramatic progress of radio technology,
effort is being made to understand whether these features are
really associated with the Galactic center or just seen along the
line-of-sight in that direction, and to determine the relative
positions of them along the line-of-sight, i.e., the
three-dimensional (3-D) spatial structure of the Galactic center.

Observations of 327~MHz absorption toward Sgr A West definitely
place Sgr A East behind Sgr A West (\citealt*{yus87}; see also
\citealt*{ped89}). \citet*{mez89} observed ring-shaped 1.3~mm
dust emission surrounding Sgr A East across the $50~\kms$ cloud
and the $20~\kms$ cloud, and suggested that Sgr A East has
expanded into these molecular clouds. Based on these
observational arguments, \citet{mez89} proposed a 3-D structure of
the Sgr A complex and concluded that the event which created Sgr
A East and the associated dust shell did not occur deep within
the GMCs but close to their surfaces facing the sun.

However, \citet*{geb89} found CO absorption toward a few Galactic
center infrared (IR) sources and suggested that the $20~\kms$
cloud may be located in front of Sgr A West. They also found some
evidence that the $50~\kms$ cloud lies partly in front of Sgr A
West.

Based on the NH$_3$ morphology and kinematics observed using the
Very Large Array (VLA), \citet*{coi99,coi00} located the Galactic
nucleus (defined to include Sgr A*, Sgr A West, and the CND
throughout this paper) behind the southern streamer (and the
$20~\kms$ cloud; see also \citealt*{gus80}) and the $50~\kms$
cloud (or the northern part of the molecular ridge) slightly
behind Sgr A East. They also argued that the distance between Sgr
A East and the $20~\kms$ cloud along the line-of-sight should be
smaller than 8.4 pc, which is the size of the SNR G~359.92-0.09
in 20~cm radio continuum images \citep*{yus87,ped89}.

\citet*{her05} updated and modified the 3-D model of \citet{coi00}
based on their additional NH$_3$ line data and more recently
published results (\citealt*{mae02} and references therein;
\citealt{par04}), as follows.  The nuclear region is placed just
inside the leading edge of Sgr A East. Only some part of the
$50~\kms$ cloud is located in front of the nucleus. The western
streamer seen in NH$_3$ emission is highly inclined to the
line-of-sight and is expanding outward with Sgr A East.  The
northern ridge is placed along the northern edge of Sgr A East and
is expanding perpendicular to the line-of-sight.  The southern
streamer passes over the nucleus in projection but probably does
not interact with it.

Together, the 3-D models above agree on the following features:

\begin{enumerate}
  \item The Galactic nucleus lies in front of Sgr A East
    but behind the southern streamer and part of the $20~\kms$ cloud along the line-of-sight.
  \item Sgr A East is expanding into the $50~\kms$ cloud,
    the northern ridge, and the western streamer.
  \item SNR G~359.92-0.09 is colliding with the southern part of the molecular ridge,
    the eastern edge of the $20~\kms$ cloud, and the southern edge of Sgr A East.
\end{enumerate}

On the other hand, contradictions among the models raise the
following questions.

\begin{enumerate}
  \item Is the nucleus in contact with or contained within Sgr A East?
  \item Is the southern streamer falling into the nucleus?
  \item Has Sgr A East expanded into the $50~\kms$ cloud significantly, or just started to contact it?
  \item Is Sgr A East colliding with the northern part of the molecular ridge?
  \item Is Sgr A East in contact with the $20~\kms$ cloud?
  \item Is the $20~\kms$ cloud located only in front of Sgr A East,
    or also extended further to the backside of it along the line-of-sight?
\end{enumerate}

It should be noted that the models above are all based on indirect
evidence like morphology, kinematics of molecular clouds, or
absorption of background radiation by these clouds, rather than on
direct, physical interactions between the objects.  To answer some
of the above questions directly, we have observed molecular
hydrogen (H$_2$) emission and constructed a 3-D picture of the
Galactic center. H$_2$ emission is an excellent tracer of
interactions between dense molecular clouds and other hot and
powerful objects, like Sgr A East.

In this paper we report observations of H$_2$ line emission from
regions of interaction between Sgr A East and other gaseous
components within the central 10 pc.  Our observations were almost
entirely of the H$_2$ 1-0 S(1) line. Unlike most previous work, we
observed this line at sufficiently high spectral resolution to
resolve the velocity profiles and at high enough angular
resolution to obtain detailed information on the spatial structure
of the emission. We also obtained measurements of the H$_2$ 2-1
S(1) line at 2.2477~$\mu$m at one location in order to investigate
the excitation mechanism of the H$_2$.

We describe the observations in Section~\ref{observation} and the
reduction of the spectroscopic data in Section~\ref{reduction}.
Based on the directions of the shocks derived from the direct
comparison of radial velocities with those from the NH$_3$(3,3)
data of \citet{mcg01}, we construct a 3-D model for the structure
of the central 10~pc in Section~\ref{structure}. In a forthcoming
paper we discuss the properties of the shocks and estimate the
explosion energy and age of Sgr A East, from which we constrain
its origin.

\section{OBSERVATIONS} \label{observation}

We surveyed four different fields in the $\hoz$ line, near the
edges of Sgr A East where interactions between its hot, expanding
gas and molecular clouds in the central 10~pc are expected (see
Fig.~\ref{field_position}). The northeastern field (hereafter
field NE) includes part of the $50~\kms$ GMC and the northern
ridge. The eastern field (field E) includes the northern portion
of the molecular ridge. The southern field (field S) extends along
the southern streamer, the northern half of which overlaps Sgr A
East; 1720~MHz OH maser emission (an indicator of shock
interactions) is also found in this field \citep{yus99a}. The
western field (field W) includes portions of the northwestern
part of the CND and the northern part of the western streamer.

The data were obtained at the 3.8~m United Kingdom Infrared
Telescope (UKIRT) during 2001 and 2003 using the facility
instrument Cooled Grating Spectrometer 4 (CGS4; \citealt*{mou90})
with its 31~l/mm echelle, 300~mm focal length camera, and a
two-pixel-wide slit. The pixel scale along the slit was
0.90~arcsec with the grating angle of $64\fdg691$ and the slit
width on the sky was 0.83~arcsec. The angular resolution, which
was affected by seeing and the optics of the spectrometer, was
about 2~arcsec ($\sim 0.08$~pc at the distance to the Galactic
center) based on the the measured FWHM of the flux profile of the
standard star along the slit. The instrumental resolution,
measured from Gaussian fits to telluric OH lines in our raw data,
was $\sim 18~\kms$. The slit length was $\sim$~90 arcsec, which is
longer than the typical size of a molecular clump of $30\arcsec$
or 1.2~pc (see the NH$_3$ map in Figure~\ref{field_position}).
CGS4 is a unique instrument in using such a long slit together
with an echelle grating. Thus we could employ the observing
technique of scanning large fields (similarly to the
low-dispersion observations of \citealt{bur92,lee05a}) with the
high spectral resolution.

Rectangular fields were observed by stepping the telescope by
3~arcsec perpendicular to the slit axis. The telescope was nodded
between object and blank sky positions every 20 minutes (one cycle
for observing a single slit position consisted of one sky exposure
and five object exposures), to allow subtraction of the background
and telluric OH line emission. The sky positions were offset by
about 2\,\fdg5 ($\Delta \alpha = -2\,\fdg03, ~ \Delta \delta =
0\,\fdg85$) from the on-source positions. We designate each slit
position `slit ' + [scanning direction] + [separation from a `base
position' in the scanning block it belongs to; in arcsec]. The
base position is a starting point of scanning each field block
and its coordinates are given in Table~\ref{h2_observations}. For
example, `slit NW12' in field NE-1 is separated from the base
position (called `slit 00') by 12~arcsec toward northwest. The
integration time at each slit position was 1000 seconds. Two
stars, HR~6496 (before 2003 May 28) and HR~6310 (on and after
that date), were observed for flux-calibration. The observations
are summarized in Table~\ref{h2_observations}.

The measurements on 2001 August 4 were performed slightly
differently from those on other nights. In order to remove bad
pixels more efficiently, the observing positions were jittered
along the slit during the 5 exposures taken on each object
($\Delta$p = 0, +1, +2, -1, and -2 pixel in sequence).

\section{DATA REDUCTION} \label{reduction}

The data were reduced in three stages.  In the first stage,
performed at the telescope by the UKIRT pipeline reduction
software {\sc ORAC-DR}, the individual frames were flatfielded and
approximately wavelength-calibrated. The next stage involved the
use of standard {\sc IRAF} \footnote{IRAF is distributed by the
National Optical Astronomy Observatories, which are operated by
the Association of Universities for Research in Astronomy, Inc.,
under cooperative agreement with the National Science Foundation.}
routines to perform sky subtraction using the sky frames,
interpolate over bad pixels, remove S-distortions and wavelength
distortion and impose an accurate (to $\pm 1~\kms$) wavelength
calibration using OH lines. We also removed stellar continua and
residual skylines, and flux-calibrated using the spectra of
HR~6496 and HR~6310. We determined the slit losses for these stars
by assuming a circularly symmetric point-spread-function (PSF)
based on the flux profile along the slit length, to estimate the
missing stellar flux. The correction factor, which varies with the
seeing, ranged from 2.06 to 2.94 \citep{lee06}.

The final stage of data reduction involved the use of {\sc MIRIAD}
\citep*{sau95,hof96}, a program package generally used for
reduction and image analysis of radio interferometric data.
However, {\sc MIRIAD} can also be used for a general reduction of
continuum and spectral line data. We employed {\sc MIRIAD} to
stack the 2-D spectral images into a single 3-D data cube for
each of the 11 fields (see Table~\ref{h2_observations}) and then
combined them into a total data cube that contains coordinate
information for every position along every slit for each slit
orientation. For more details see \citet{lee05b}.

The resulting integrated intensity map for the combined data cube
is shown in Fig.~\ref{cube_all}. A smoothed version of the total
cube, produced by convolving with a Gaussian profile of FWHM =
$3\arcsec$ to give a higher signal-to-noise (S/N) ratio, is also
presented in Fig.~\ref{cube_all_smooth3}.

\section{H$_2$ EMISSION AROUND SGR A EAST} \label{h2emission}

H$_2$ is the most abundant molecule in the interstellar medium
(ISM). We cannot, however, observe H$_2$ directly in cold, dense
molecular clouds because the lowest energy levels of H$_2$ are too
high to be excited in these environments ($T <$ 50~K). Instead,
H$_2$ emission is observed in more active regions, for example, in
warm regions heated by shocks or in the surfaces of clouds
illuminated by far-ultraviolet (far-UV) radiation. H$_2$ emission
has been found associated with star-forming regions, SNRs,
planetary nebulae, and active galactic nuclei.

Since the first detection by \citet*{gat84}, a number of groups
have observed the Galactic center in H$_2$ line emission.  Using a
Fabry-Perot (FP) etalon
\citet*{gat86} mapped the CND in the $2.1218~\micron$ $\hoz$
emission with an angular resolution of $18\arcsec$. They found
that the CND
has a broken, clumpy appearance. \citet*{bur92} obtained images in
various emission lines (He\,I $2.058~\micron$, Br$\gamma$
$2.166~\micron$, and $\hoz$) by scanning the telescope
perpendicular to the slit, covering spatially an area of
$103\arcsec \times 145\arcsec$ and spectrally the entire K window
(2.0--2.4~$\micron$) with a resolution of $\lambda / \Delta\lambda
= 400$. The near-IR images show a cluster of He emission line
stars, the mini-spirals in Br$\gamma$, and the CND in H$_2$. They
also observed the H$_2$ emission peak in the CND with normal
spectroscopic techniques, and suggested collisional fluorescence
as the emission mechanism. On larger scales,
\citet*{pak96a,pak96b} surveyed the Galactic plane in the $\hoz$
emission along a 400-pc strip.  They found that H$_2$ emission can
be seen throughout the surveyed region, peaking toward Sgr A. They
also mapped the central 50~pc with a beam size of $3\farcm3$ in
diameter. \citet*{war99} detected the $\hoz$, $\htos$, and $\rm
1\!-\!0 ~ S(0)$ lines at the position `A' of the 1720~MHz OH maser
detection in Figure~\ref{field_position}. But they could not
constrain the H$_2$ excitation mechanisms using line ratios
because of large uncertainties in the line fluxes.
\citet*{yus99b,yus01} also surveyed H$_2$ line emission around the
regions where the OH masers have been detected, and imaged the CND
and most of the Sgr A East region using NICMOS on the {\it Hubble
Space Telescope (HST)}. Their H$_2$ image has the highest spatial
resolution ($0\farcs20$) obtained so far. To study the kinematics
of the CND, they also observed this field using a FP etalon, with
a FWHM spectral resolution of $\sim 75~\kms$. Based on their
results, combined with the OH detection, they suggested that the
H$_2$ gas is shocked and accelerated by the expansion of Sgr A
East into the $50~\kms$ cloud and the CND.

Over the past two decades, in spite of dramatic advances in H$_2$
observations of the central 10~pc, there are still many remaining,
unsolved questions.  The H$_2$ excitation mechanism is still in
debate (UV-heated vs. shock-heated), particularly between studies
using line ratios \citep{pak96a,pak96b} and those based on the
relationships with the OH masers \citep{war99,yus99b,yus01}. The
previous observations were concentrated on the CND, the brightest
H$_2$ feature, rather than the regions where interactions between
Sgr A East and the surrounding molecular clouds are expected.
Also, although the spatial resolution has greatly increased, the
spectral resolution has not been high enough to investigate the
detailed kinematics (e.g., line profiles) in the interaction
regions. Considering that the widths of observed H$_2$ lines are
generally less than $45~\kms$, which is the critical velocity for
shock dissociation in molecular clouds (e.g. \citealt*{smi91}),
the spectral resolutions (75--130~$\kms$) of previous
spectroscopic observations with FP etalons
\citep*{gat86,yus99b,yus01} are too low. To study the spatial and
dynamical relationships between the various components in the
central 10~pc, it is necessary to observe additional interaction
regions other than the CND, in the H$_2$ emission at high spatial
and spectral resolutions.

\subsection{Projected Morphology of Sgr A East in the H$_2$ Emission}

The morphology of Sgr A East has frequently been determined from
the maps of 6~cm synchrotron radiation
(\citealt*{eke83,yus87,ped89}; see Figure~\ref{field_position}).
We investigate the morphology of the Sgr A East boundary in the
H$_2$ emission, which is imaged for the first time in this study.
Figure~\ref{h2_boundary} shows our model of the Sgr A East
boundary in projection based on the H$_2$ intensity map. The
H$_2$ emission is significantly stronger than the RMS noise of $5
\times 10^{-21}~\intenwat$ (the blue color represents roughly a
4-$\sigma$ detection) in most of the fields observed. The intense
H$_2$ emission arises from the interaction regions related to the
$50~\kms$ cloud and the northern ridge in the northeastern field
and from the regions related to the CND and the western streamer
in the western field. In the eastern and southern fields, the
H$_2$ emission is weaker.

An elliptical boundary is defined to trace the outer edges of the
H$_2$ emitting regions with the center at $\rm \alpha = 17^h 45^m
42\fs13, ~ \delta = -29\degr0\arcmin8\farcs6$ (J2000), which is
offset from Sgr A* by ($+32\arcsec$, $+18\arcsec$) or
$\sim1.5$~pc at the distance of 8.0~kpc to the Galactic center
\citep*{rei93}. The ellipse has a semi-major radius of $a =
135\arcsec$ ($= 2\farcm25 = 5.4$~pc), a semi-minor radius of $b =
95\arcsec$ ($= 1\farcm58 = 3.8$~pc), and a position angle of
$30\degr$ from north to east, which is almost parallel to the
Galactic plane whose position angle is $\simeq 34\degr$. The
elliptical boundary is determined qualitatively so that it can
include all H$_2$ emission brighter than 10-$\sigma$ and be kept
consistent in shape with the morphology in 6-cm continuum
\citep{yus87} and dust map \citep{mez89}. This model of H$_2$
boundary might be revised by other studies since the model may
depend on geometry or cloud structure as the H$_2$ emission arises
only from local interface regions between Sgr A East and the
clouds and our H$_2$ survey does not cover the whole region around
Sgr A East.

The only conflict with this model is the southern field where the
H$_2$ emission is situated well inside of the synchrotron shell in
projection (compare Figure~\ref{field_position} and
Figure~\ref{h2_boundary}). We suggest that this southern H$_2$
emission is not radiated from the southern-most edge of the Sgr A
East shell but from a position where the tilted surface of the
shell contacts a molecular cloud (i.e. the southern streamer) in
front of or behind it. Alternatively, the H$_2$ emission may be
extended more to the south from the detected position but severely
diminished due to a very high extinction toward the southern part
of the southern streamer (see the marginally detected H$_2$
emission in Figure~\ref{pvd5} where all the data in the southern
field are summed to increase the S/N ratio). Dust emission is
strong in this direction (\citealt*{zyl98}; see Figure~9 of
\citealt{mcg01}) and the NH$_3$ opacity in this region is much
higher ($\tau_{NH_3~(1,1)}$ = 2--5) than in the region where
H$_2$ is detected ($\tau_{NH_3~(1,1)} << 1$; see Figure~2 of
\citealt*{her05}). The 1720~MHz OH maser detected at several
positions around this region (see Figure~\ref{field_position}) by
\citet*{yus96,yus99a} may support this hypothesis. Those authors
interpreted the maser detections as indicators of shocks from Sgr
A East toward its nearby molecular cloud, although \citet*{coi00}
and \citet*{her05} argued that they originate from the
interaction between Sgr A East and the SNR G~359.92-0.09. A
similar interpretation is also possible for the $50~\kms$ cloud.
In the northeastern field of our H$_2$ observation, the H$_2$
intensity decreases toward the center of the $50~\kms$ cloud
(Figure~\ref{h2_boundary}). In this direction the dust emission
is the strongest in the central 10~pc and the NH$_3$ opacity is
as high as in the southern streamer \citep*{her05}. Thus it is
possible that, even though Sgr A East has expanded deeply into
this cloud, the shock-excited H$_2$ emission is highly obscured.

Assuming the same center and the same position angle as our H$_2$
boundary, the projected 6~cm continuum shell can be simplified
with an ellipse with $a_{6cm} = 1\farcm7 = 4.2$~pc and $b_{6cm} =
1\farcm3 = 3.0$~pc (Figure~\ref{other_boundary}). These dimensions
are smaller than those of the H$_2$ boundary by about 20 per cent.
The boundary of Sgr A East defined by H$_2$ emission is more
consistent with the dust ring observed by \citet*{mez89} than
with the outer edge of the 6~cm shell. The partial ring of 1.3~mm
dust emission surrounds the 6~cm synchrotron emission (see
Figure~3b of \citealt*{mez89}). This dust ring is well followed
by the molecular clouds seen in NH$_3$ (see
Figure~\ref{other_boundary}). The $50~\kms$ cloud, the northern
ridge, the western streamer, and the southern streamer in NH$_3$
emission are easily matched with the dust ridges. Assuming the
same center and the same position angle with our H$_2$ boundary,
the dust ring can be represented with an ellipse with $a_{dust} =
2\farcm5 = 6.0$~pc and $b_{dust} = 1\farcm5 = 3.7$~pc, which is
nearly identical with the H$_2$ ellipse although the major axis
of the dust ring is slightly (about 10 per cent) longer.

In a comparison between their 1.3~mm map and the 6~cm map,
\citet*{mez89} argued that the magnetic field of the synchrotron
radiation is created in regions of the shell well down-stream of
the shock front.  This argument, together with the fact that the
dust ring coincides well with the outer boundary of the H$_2$
emission, implies that the H$_2$ boundary defined here actually
traces the shock front of Sgr A East.

\subsection{Shock Interactions between Sgr A East and Molecular
Clouds} \label{shock}

\citet{lee03} observed H$_2$ emission lines from interfaces
between Sgr A East and each of the $50~\kms$ cloud (the GMC
M-0.02-0.07) and the northern ridge. Based on the kinematics,
they concluded that the H$_2$ molecules are excited by shocks
from Sgr A East although fluorescence works partially as well.

In this paper, large-scale spatial and kinematic structure of Sgr
A East and the molecular clouds is investigated by comparing the
position-velocity diagrams (PVDs) of the H$_2$ and NH$_3$
emission. The H$_2$ emission traces hot ($\sim 2000$ K) gas and
the NH$_3$ cool ($\la~100$ K) gas. In Figure~\ref{pvd_position}
we superimpose the axes of the six PVDs shown in
Figures~\ref{pvd1}--\ref{pvd6}. Each PV cut is selected to cover
bright regions in both H$_2$ and NH$_3$ in general. More
specifically, cut C1 is designated to pass through three H$_2$
emission peaks related to the $50~\kms$ cloud and a H$_2$
emission feature around the northern end of the molecular ridge.
Cut C2 is approximately along the northern ridge which is curved
toward the CND. To compare the $50~\kms$ cloud and the northern
ridge, cut C3 is laid across the two clouds and their related
H$_2$ emission. Cut C4 goes along the whole length of the
molecular ridge and the $50~\kms$ cloud. The purpose of this cut
is to study the relationships between these two clouds and the
H$_2$ emission in this region. Similarly, cut C5 is made to pass
through both the CND and the southern streamer in order to
investigate if the streamer is just a foreground feature as
suggested by \citet{her05}. Finally, cut C6 goes across both the
CND and the western streamer. This cut also covers a small patch
of H$_2$ emission in the westernmost part (find more discussions
on this H$_2$ feature in section~\ref{shock_directions}).

In Figure~\ref{pvd1}, most of the NH$_3$ emission contours trace
the $50~\kms$ cloud and extend to the northern end of the
molecular ridge (at positions $< -100\arcsec$). The small patch
of emission at positions $10\arcsec$--$40\arcsec$ and velocity of
about $0~\kms$ corresponds to the northern end of the northern
ridge. The H$_2$ emission observed in our northeastern field (at
positions between $-50\arcsec$ and $50\arcsec$) shows similar
velocity peaks as NH$_3$. The velocity extension of H$_2$ is much
broader; by as much as $60~\kms$. This implies that strong shocks
are propagating into the $50~\kms$ cloud and the H$_2$ emission
arises from turbulent post-shock gas.

In the PVD for cut C4 (Figure~\ref{pvd4}) which passes through
the center of the molecular ridge along its length, we can see
that the H$_2$ contours are broader in velocity and extend
farther to the red side than NH$_3$. Thus we are certain that Sgr
A East is in physical contact with and driving shocks into the
molecular ridge too, at least into its northern part.

Cut C5 follows the southern streamer (see Figure~\ref{pvd5}). The
NH$_3$ at $\sim +30~\kms$ starts from $-60\arcsec$, continues
through the nuclear region (between $+90\arcsec$ and
$+140\arcsec$), and reaches beyond the northern boundary of the
CND. The weak NH$_3$ features at both sides of the southern
streamer (at 10, 60, and 80~$\kms$) are the satellite hyperfine
lines of the strong main line \citep{mcg02,her05}, so have no
additional kinematic meaning. The NH$_3$ feature with a very high
velocity gradient between $+40\arcsec$ and $+100\arcsec$ is
thought to be associated with the CND. In Figure~\ref{pvd5}, all
cuts parallel to C5 in field S are combined to increase the S/N
ratio. The H$_2$ emission is clearly blue-shifted with respect to
NH$_3$ by at least $20~\kms$ for both clouds. Thus the Sgr A East
shock is suspected to be the accelerator of the H$_2$ molecules
also in the southern streamer and the CND.

In Figure~\ref{pvd6}, the northwestern part of the CND is shown at
positions $> 20\arcsec$. The NH$_3$ emission from this cloud has
a very broad velocity distribution ($\sim 100~\kms$) reflecting
very complicated and energetic gas motions in the nuclear region.
The related H$_2$ contours from $20\arcsec$ to $30\arcsec$ are as
wide in velocity. The NH$_3$ emission contours here peak at $\sim
80~\kms$ and are skewed toward positive velocities while the
H$_2$ emission is peaked at $\sim 50~\kms$ and is also bright
toward lower velocities. This indicates that this part of the CND
is located in front of Sgr A East and being pushed toward us by
its expansion.

The second molecular feature in Figure 12 is the northern half of
the western streamer at $-30\arcsec$ to $10\arcsec$. The H$_2$
contours from the shocked gas at $\sim 0\arcsec$ are as wide as
$\sim 130~\kms$ and extended slightly farther (by $\sim 30~\kms$)
both to the blue side and the red side than the NH$_3$. Thus it
is likely that this part of the western streamer actually
surrounds the western part of Sgr A East and is being swept up by
both the front and back of the expanding shell.

In summary, throughout the region, the H$_2$ emission lines are
significantly broader than the NH$_3$ emission lines by a factor
of 2--3. The H$_2$ profiles extend either blue-ward or red-ward
of the NH$_3$ which traces the systemic (rest) velocity of each
molecular cloud more accurately. This kinematics implies that the
H$_2$ emission around Sgr A East clearly originates from shocks
propagating into the molecular clouds (see \citealt{lee03} for
more detailed discussion on the H$_2$ excitation including
partial role of non-thermal excitation).

\section{THREE-DIMENSIONAL SPATIAL AND KINEMATIC STRUCTURE OF THE CENTRAL 10 PARSECS} \label{structure}

In this section we will discuss the features within the Galactic
center region.  Using our data and past observations (described
below), we develop a 3-D view of the region, which we show in
Figure~\ref{3D_model}. Justification for the placement of the
various components is given in the following sub-sections.

\subsection{Sgr A East as a Key Object in Understanding the 3-D Structure}

Sgr A East surrounds the Sgr A* complex (including Sgr A West and
the CND) in projection (see Figure~\ref{field_position}). Along
the line of sight, absorption of non-thermal radiation is obvious
evidence that the Sgr A* complex lies in front of the Sgr A East
shell \citep*{yus87,ped89}. A number of arguments, however,
suggest that Sgr A* is in physical contact with or possibly
embedded within the hot cavity of the Sgr A East shell (see
\citealt*{mor96,yus00,mae02} and references therein). For example,
there exists faint non-thermal emission detected at 90~cm toward
the thermally ionized gas although most of the non-thermal
emission from Sgr A East is absorbed by the ionized gas associated
with Sgr A West. This may indicate that Sgr A West is embedded in
Sgr A East and the detected radiation is from the region between
Sgr A West and the front-most edge of the Sgr A East shell toward
us \citep{yus00}.

There is also observational support for the argument that Sgr A
East is in physical contact with and driving shocks into the CND.
\citet*{yus99b} found a linear filament of H$_2$ emission located
at the western edge of the CND running parallel to the Sgr A East
shell. This H$_2$ feature is thought to occur by shock-heating,
as indicated by its morphology, the association with a source of
1720~MHz OH maser, and the lack of evidence for UV heating in the
form of thermal radio continuum or Br$\gamma$ emission. In
addition, a north-south ridge outlining the eastern half of the
CND can be seen in the 20~cm continuum emission \citep{yus00}.
This elongated ridge is also detected at 90~cm
\citep*{ped89,yus99b}, suggesting that it is a non-thermal feature
related to Sgr A East. As we will see in
section~\ref{shock_directions}, Sgr A East is driving shocks into
the northwestern part of the CND and accelerating the H$_2$ gas
toward negative velocities. This interpretation supports the
argument of \citet{yus00} that the H$_2$ filament detected along
the western edge of the CND is shock-heated. Absorption features
in H$_2$CO, OH, HI, and HCO$^+$ spectra with highly negative
radial velocities ($V_{LSR} \simeq -190~\kms$) have also been
observed toward Sgr A West
\citep*{mar92,pau93,yus93,yus95,zha95}. The kinematics and spatial
distribution of this gas place it at the Galactic center and
\citet{yus00} interpret its highly negative velocity as a result
of acceleration by Sgr A East. Thus we conclude that Sgr A East
is situated within the central 10~pc and that it is physically
interacting with the Sgr A* complex.

Sgr A East is also actively interacting with the molecular clouds
in the central 10 pc \citep{mcg01,lee03,her05}. In
section~\ref{shock} we demonstrate that Sgr A East is driving
shocks into the surrounding clouds. As a result, we can determine
the relative locations of Sgr A East and the clouds in the line of
sight, based on the relative radial velocities between the
shocked and unshocked gas. Hence Sgr A East is used as a key
object in understanding the 3-D structure around the nucleus of
our Galaxy.

\subsection{Radial Velocities in H$_2$ and NH$_3$ - Shock Directions and Spatial
Relationships} \label{shock_directions}

In Section~\ref{shock}, Sgr A East is shown to be physically in
contact with and driving shocks into all of the surrounding
molecular clouds; the $50~\kms$ cloud, the northern ridge, the
molecular ridge, the southern streamer, the northwestern part of
the CND, and the western streamer. In this section, we compare
the velocity structures of the mutually related H$_2$ and NH$_3$
emission using the PVDs (Figures~\ref{pvd1}--\ref{pvd6}) in order
to determine shock directions and positional relationships along
the line-of-sight between Sgr A East and the molecular clouds.

In Figure~\ref{pvd1}, the H$_2$ emission shows much broader
velocity extension than the NH$_3$ contours of the $50~\kms$
cloud, both to positive and negative velocities. This implies
that strong shocks are propagating both toward us and in the
opposite direction along the line-of-sight, within the $50~\kms$
cloud. We can understand this if the western portion of the
$50~\kms$ GMC actually envelops Sgr A East, which is expanding
into the cloud at both its front and back surfaces.

On the other hand, in Figure~\ref{pvd1}, the H$_2$ emission from
the northern end of the molecular ridge has a narrow and very
similar velocity distribution to the NH$_3$ emission. However, in
Figure~\ref{pvd4} for cut C4, which passes through the molecular
ridge, the H$_2$ contours are broader in velocity and extend
farther to the red side than NH$_3$. The velocity shift of H$_2$
here is about $20~\kms$ which is much smaller than in the
$50~\kms$ cloud. However, considering that the molecular ridge is
located at the outermost edge of the Sgr A East boundary and that
its projected width is only about 1~pc, shocks from Sgr A East
would propagate into the cloud nearly perpendicular to the
line-of-sight and consequently the radial component of velocity
shift of the shocked gas must be small. Nevertheless, we expect
that the northern end of the molecular ridge is tipped slightly
to the backside of Sgr A East since the H$_2$ emission there is
red-shifted, i.e. the hot cavity of Sgr A East is located in
front of the ridge and pushing its material farther away from us.

As for the small emission patch of the northern ridge in
Figure~\ref{pvd1}, it is difficult to distinguish between the
H$_2$ emission that originated from this cloud and that from the
$50~\kms$ cloud since two molecular clouds overlap along the
line-of-sight. This problem is similar to the other PVDs related
to the northern ridge (Figures~\ref{pvd2}~\&~\ref{pvd3} for cuts
C2 and C3, respectively). However, there is a common aspect in
these PVDs; there is no H$_2$ emission more blue-shifted than the
NH$_3$. This implies either that Sgr A East is located in front of
the northern ridge along the line-of-sight or that the H$_2$
emission does not originate from shocked gas. The latter
interpretation is not likely as seen in Section~\ref{shock}; in
Figure~\ref{pvd2}, the positions of the bright peaks of broad (as
much as $100~\kms$) H$_2$ line emission are more closely
coincident with two NH$_3$ peaks of the northern ridge (at $\sim
0~\kms$) than with the single peak of the $50~\kms$ cloud.
Evidence for shocked H$_2$ emission in the northern ridge can
also be found in \citet{lee03}. Therefore we conclude that the
northern ridge is located to the far-side of Sgr A East and is
being accelerated away from us.

In Figure~\ref{pvd5} for cut C5, H$_2$ emission is only seen
bright at $\sim 50\arcsec$ but at two separate velocities of the
southern streamer ($\sim +30~\kms$) and the CND (from $\sim
-70~\kms$ to $\sim -10~\kms$). The H$_2$ emission is clearly
blue-shifted with respect to NH$_3$ by at least $20~\kms$ for
both clouds. Hence we conclude that Sgr A East is located behind
the southern streamer and the southern part of the CND,
respectively.

Figure~\ref{pvd6} includes emission from three different molecular
features. One is the northwestern part of the CND at positions $>
20\arcsec$, where the NH$_3$ contours peak at $\sim 80~\kms$. The
related H$_2$ contours peak at $\sim 50~\kms$ and are broadened
toward lower velocities. This indicates that this part of the CND
is located in front of Sgr A East and being pushed toward us by
its expansion. The second feature at $-30\arcsec$ to $10\arcsec$
is the northern half of the western streamer. The related H$_2$
contours at $\sim 0\arcsec$ are extended slightly farther (by
$\sim 30~\kms$) both to the blue side and the red side than the
NH$_3$. We interpret this as this part of the western streamer
partially surrounds the western part of Sgr A East and is being
swept up by both the front and back of the expanding shell. The
third feature is the H$_2$ emission at $\sim -35\arcsec$, which
must certainly originate in shocked gas considering its wide
velocity distribution of $\sim 80~\kms$. It is not clear, however,
from which molecular cloud it arises. Since its position is beyond
the boundary of Sgr A East
(Figures~\ref{h2_boundary}~\&~\ref{pvd_position}), we cannot
determine its line-of-sight position with respect to Sgr A East.
This H$_2$ feature beyond Sgr A East may be associated with
another phenomenon or accelerating source. For example, it might
be a manifestation of the bipolar streamers or outflows from the
Galactic nucleus which are suggested by radio and X-ray
observations \citep*{yus87,mae02}. Or it might be shocked gas
which has leaked out from the bubble (see the radio continuum
feature toward northwest in Figure~\ref{field_position}). Further
studies are needed to understand the nature and origin of this
H$_2$ feature.

\subsection{Three-Dimensional Spatial Structure of the Central 10 Parsecs}

In the previous sections, we presented a 3-D model of the inner
10 pc of the Galaxy. Our model agrees with the previous studies
\citep{mez89,coi00,her05} on the following points.
\begin{enumerate}
  \item The Galactic nucleus lies in front of Sgr A East
    and behind the southern streamer and a part of the $20~\kms$ cloud along the line-of-sight.
  \item Sgr A East is expanding into the $50~\kms$ cloud (M-0.02-0.07),
    the northern ridge, and the western streamer.
\end{enumerate}

In addition to the above, we suggest the following.
\begin{enumerate}
  \item Sgr A East is expanding deeply into the western edge of the $50~\kms$ GMC which
    envelops it both at the front and rear of the ionized shell.
  \item The molecular ridge is approximately at the same distance as the center of Sgr A
    East, but the northern end of the ridge is tilted slightly to the back of it.
  \item The northern ridge is in contact with the backside of Sgr A East.
  \item The northern-most end of the southern streamer and the CND lie in front of Sgr A East
    and are being pushed toward us by it.
  \item The northern part of the western streamer is located at the same distance as
    the center of Sgr A East and barely envelops the western edge of it.
\end{enumerate}

Based on the outer boundary of the Sgr A East cavity defined by
the H$_2$ line emission and above conclusions, we suggest a
revised model for the 3-D structure of the central 10~pc as shown
in Figure~\ref{3D_model}.

\section{CONCLUSIONS}

Based on the H$_2$ emission map, we determine the outer boundary
of Sgr A East where it is driving shocks into the surrounding
molecular clouds, to be approximately an ellipse with the center
at ($+32\arcsec$, $+18\arcsec$) or $\sim1.5$~pc offset from Sgr
A*, a major axis of length 10.8~pc, which is nearly parallel to
the Galactic plane, and a minor axis of length 7.6~pc. This
boundary is significantly larger than the synchrotron emission
shell \citep*{eke83,yus87,ped89} but is closely consistent with
the dust ring suggested by \citet*{mez89}.

Since Sgr A East is in physical contact with all of its nearby
molecular clouds (the $50~\kms$ cloud, the northern ridge, the
molecular ridge, the southern streamer, the CND, and the western
streamer),  we are able to determine the positional relationships
between Sgr A East and the molecular clouds along the
line-of-sight using the shock directions as indicators.  Based on
the determined relationships and the strong evidence that Sgr A
East is in contact with the nucleus, we suggest a revised model
for the 3-D spatial structure of the central 10 pc of our Galaxy
modifying the previous models of \citet*{mez89}, \citet*{coi00},
and \citet*{her05}.

Our conclusions on the 3-D structure resolve most of the debates
in previous studies as follows:
\begin{enumerate}
  \item Is the nucleus in contact with or contained within Sgr A East?
    -- The Galactic nucleus is in physical contact with Sgr A East
    since the CND is pushed toward us by the expanding hot cavity of Sgr A East.
  \item Is the southern streamer falling into the nucleus?
    -- It is highly probable that the southern streamer is falling into the nuclear
    region and feeding the CND. Sgr A East is driving shocks into the northern-most part of
    this cloud where it meets the CND in projection.
  \item Has Sgr A East expanded into the $50~\kms$ cloud significantly, or just started
    to contact it?
    -- In the H$_2$ data of the northeastern field, we can see that most of this region
    is filled with shocked gas from the $50~\kms$ cloud. The area corresponds to
    at least one third of that of the entire cloud. Thus Sgr A East
    has significantly penetrated the cloud.
    If the cloud is wrapping around the Sgr A East shell with a casual morphological
    coincidence, however, the shocked layer might still be thin and on the surface.
  \item Is Sgr A East colliding with the northern part of the molecular ridge?
    -- Yes, we detected shocked H$_2$ emission from the northern-most part of this cloud.
\end{enumerate}

However, we cannot answer the questions related to the $20~\kms$
cloud. We know the branches (the southern streamer and the western
streamer) from this GMC are interacting with Sgr A East but the
main body of this cloud is located far to the south, beyond the
scope of our observations. Therefore the position and extent
along the line-of-sight of this GMC in Figure~\ref{3D_model} is
uncertain. According to \citet*{coi99,coi00}, SNR G~359.92-0.09
is expanding into the molecular ridge, the $20~\kms$ cloud, and
Sgr A East. Thus, if we observe the H$_2$ emission around SNR
G~359.92-0.09 in a similar way to the 3-D observations for Sgr A
East, the questions about the $20~\kms$ cloud could also be
answered.

\acknowledgments

We thank the staff at UKIRT for their excellent support during our
successful observations. Kind answers from Andy Adamson, Paul
Hirst, and Tom Kerr concerning CGS4 were critically helpful for
this work.  Fig.~\ref{field_position} is reproduced from Fig.~10
of \citet{mcg01} by permission of the AAS. The United Kingdom
Infrared Telescope is operated by the Joint Astronomy Centre on
behalf of the U.K. Particle Physics and Astronomy Council. TRG's
research is supported by the Gemini Observatory, which is operated
by the Association of Universities for Research in Astronomy,
Inc., on behalf of the international Gemini partnership of
Argentina, Australia, Brazil, Canada, Chile, the United Kingdom
and the United States of America.
  This work was supported by the Korea Science and Engineering
  Foundation (KOSEF) grant funded by Korea government (MOST)
  No. R01-2005-000-10610-0.

\clearpage

\begin{figure}
  \caption{
  Schematic drawing of the 3-D structure
of the central 10 pc (the front, side, and top view in a clockwise
direction). Black dots indicate Sgr A*. Sgr A West and the CND are
simplified as two ellipses surrounding it. This model structure is
based on the results of this work. See Section~\ref{structure} for
details.}
  \label{3D_model}
\end{figure}


\begin{figure}
  \caption{
  Field positions for the H$_2$ observations.
  The color image is the 6~cm continuum map of
the central 10 pc \citep{yus87} with NH$_3$(3,3) emission contours
superimposed from \citet{mcg01}. Sgr A East (red shell) surrounds
Sgr A* (the black dot in the center) and Sgr A West (mini-spirals
in green and blue) in the radio continuum. The four fields
observed in $\hoz$ using a slit-scanning technique are indicated
by the labelled white boxes; these are the northeastern (NE;
composed of 24 parallel positions of the $90\arcsec$-length slit),
eastern (E; 10 slit positions), southern (S; 10 slit positions)
and western (W; 10 slit positions) fields. Each field is divided
into 2 or 4 scanning blocks (labelled with numbers) by black solid
lines. The narrow box (labelled `0') between field NE and field E
is a supplementary field which is composed of only two slit
positions and belongs to field E. This field (E-0) was observed to
study the relationship between Sgr A East and compact HII regions
in its eastern side, and the result will be reported separately.
Contour levels are in intervals of $4~\sigma$, where $\sigma =
0.33~\intennht$ (the beam size is $\sim 15\arcsec \times
13\arcsec$), and the scale bar ranges from 0 to 0.7~$\sintennht$
(the beam size is $3\farcs4 \times 3\farcs0$). Letters (A, B, D,
and G) mark the positions of OH (1720 MHz) masers \citep{yus99a}.
  \label{field_position}}
\end{figure}

\begin{figure}
  \caption{
  The velocity-integrated $\hoz$ map of the entire
surveyed region extracted from the combined data cube. The
color-scaled intensity level is indicated by the side bar on the
right in units of $\intenwat$. }
  \label{cube_all}
\end{figure}

\begin{figure}
  \caption{
  The velocity-integrated $\hoz$ map of the entire surveyed
region shown in Figure~\ref{cube_all}, smoothed with a
$3\arcsec$-FWHM Gaussian.  The color-scaled intensity level is
indicated by the right-side wedge in units of $\intenwat$.}
  \label{cube_all_smooth3}
\end{figure}


\begin{figure}
  \caption{
  Definition of the Sgr A East boundary in H$_2$ emission.
An ellipse defining the outer boundary of Sgr A East is overlaid
on the integrated intensity map of $\hoz$ line emission (smoothed
by Gaussian with FWHM = $5\arcsec$) with contours for NH$_3$~(3,3)
emission from \citet{mcg01}. The color-scaled intensity level is
indicated by the right-side bar in units of $\intenwat$ and the
contour levels are in intervals of 3-$\sigma$ (the RMS noise
$\sigma = 0.33~\intennht$ where the beam size is $\sim 15\arcsec
\times 13\arcsec$). The major and minor axes of the ellipse are
also indicated; the cross at the center of the image represents
the position of Sgr A*. }
  \label{h2_boundary}
\end{figure}


\begin{figure}
  \caption{
  Two morphological models based on the 6~cm continuum map
of \citet{yus87} (inner ellipse) and the dust map of
\citet{mez89} (outer ellipse) are superimposed on the gray-scaled
version of Figure~\ref{field_position}.}
  \label{other_boundary}
\end{figure}

\begin{figure}
  \caption{
  Positions of the cuts for the position-velocity diagrams
(PVDs) of $\hoz$ and NH$_3$~(3,3) emission in
Figures~\ref{pvd1}--\ref{pvd6}. For each cut, the labeled end
indicates the direction of positive offset and the small cross
corresponds to the reference position in each PVD. The NH$_3$
contours are labeled for the molecular clouds around Sgr A East.
The $\hoz$ map is smoothed by a Gaussian with FWHM = $5\arcsec$.
The intensity scale and contour levels are the same as
Figure~\ref{h2_boundary}. }
  \label{pvd_position}
\end{figure}


\begin{figure}
  \caption{
  Position-velocity
diagram for $\hoz$ and NH$_3$~(3,3) emission along cut C1. Thick
contours are for H$_2$ emission and thin contours are for NH$_3$.
The contour levels are 2, 4, 6, 8, 10, 20, 40, 60, 80, and
100-$\sigma$ for both contours where $\sigma_{H_2} = 1.5 \times
10^{-22}~\sintenwat$ and $\sigma_{NH_3} = 0.01~\sintennht$.
Positions in units of arcsec are relative to the reference
position which is marked on the cut in Figure~\ref{pvd_position}.
Thick horizontal lines indicate the boundaries of the H$_2$
fields. }
  \label{pvd1}
\end{figure}

\begin{figure}
  \caption{
  The same as Figure~\ref{pvd1} but for cut C2. }
  \label{pvd2}
\end{figure}

\begin{figure}
  \caption{The same as Figure~\ref{pvd1} but for cut C3. }
  \label{pvd3}
\end{figure}

\begin{figure}
  \caption{The same as Figure~\ref{pvd1} but for cut C4. }
  \label{pvd4}
\end{figure}

\begin{figure}
  \caption{
  The same as
Figure~\ref{pvd1} but for cut C5. All cuts parallel to C5 in
field S are summed to increase the S/N ratio. The broad H$_2$
contours at $\sim 170\arcsec$ are related to the $50~\kms$ cloud
and the northern ridge.}
  \label{pvd5}
\end{figure}

\begin{figure}
  \caption{The same as Figure~\ref{pvd1} but for cut C6. }
  \label{pvd6}
\end{figure}


\begin{table}
\begin{center}
\caption{H$_2$ Observations with CGS4 and the echelle at
UKIRT\label{h2_observations}}
\footnotesize
\begin{tabular}{lclllcc}
\tableline\tableline
Field\tablenotemark{a} & Date (UT) & \multicolumn{2}{c}{Base Position\tablenotemark{b} (J2000)}
    & Slit names\tablenotemark{c} & P.A.\tablenotemark{d} & Seeing\tablenotemark{e} \\
    & (yyyy/mm/dd) & R.A. & Dec   & (number of slits) &       & (FWHM) \\
\tableline
\multicolumn{7}{c}{$\hoz ~ (\lambda = 2.1218 \micron)$} \\
\tableline
NE-1 & 2001/08/3--4\tablenotemark{f}
                        & $\rm 17^h 45^m 45\fs95$ & $\rm -28\degr 59\arcmin 05\farcs16$ & NW15, NW12...SE12, SE15 (11) & $40\degr$ & $2.0\arcsec$ \\
NE-2 & 2001/08/04\tablenotemark{f}  & $\rm 17^h 45^m 45\fs86$ & $\rm -28\degr 59\arcmin 06\farcs54$ & SE18, SE21, SE24 (3)              & $40\degr$ & $2.1\arcsec$ \\
NE-3 & 2003/05/23     & $\rm 17^h 45^m 45\fs87$ & $\rm -28\degr 59\arcmin 04\farcs16$ & NW15, NW18...NW24, NW27 (5)       & $40\degr$ & $2.4\arcsec$ \\
NE-4 & 2003/05/28     & $\rm 17^h 45^m 45\fs95$ & $\rm -28\degr 59\arcmin 05\farcs16$ & NW30, NW33...NW39, NW42 (5)       & $40\degr$ & $1.8\arcsec$ \\
E-0  & 2001/08/04\tablenotemark{f} & $\rm 17^h 45^m 50\fs60$ & $\rm -28\degr 59\arcmin 44\farcs30$ & 00, SE03 (2)                      & $40\degr$ & $2.0\arcsec$ \\
E-1  & 2003/05/29     & $\rm 17^h 45^m 48\fs30$ & $\rm -29\degr 00\arcmin 15\farcs00$ & 00, S03...S09, S12 (5)            & $-90\degr$ & $2.1\arcsec$ \\
E-2  & 2003/06/01     & $\rm 17^h 45^m 48\fs60$ & $\rm -29\degr 00\arcmin 13\farcs80$ & S15, S18...S24, S27 (5)           & $-90\degr$ & $1.7\arcsec$ \\
S-1  & 2003/05/30     & $\rm 17^h 45^m 42\fs30$ & $\rm -29\degr 01\arcmin 53\farcs00$ & 00, W03...W12, W15 (6)            & $0\degr$ & $2.0\arcsec$ \\
S-2  & 2003/06/01     & $\rm 17^h 45^m 42\fs27$ & $\rm -29\degr 01\arcmin 50\farcs60$ & W18, W21, W24, W27 (4)            & $0\degr$ & $1.7\arcsec$ \\
W-1  & 2003/05/31     & $\rm 17^h 45^m 34\fs69$ & $\rm -29\degr 00\arcmin 06\farcs40$ & 00, NE03...NE09, NE12 (5)         & $-60\degr$ & $2.2\arcsec$ \\
W-2  & 2003/05/31     & $\rm 17^h 45^m 34\fs80$ & $\rm -29\degr 00\arcmin 05\farcs00$ & NE15, NE18...NE24, NE27 (5)       & $-60\degr$ & $2.2\arcsec$ \\
\tableline
\end{tabular}
\tablenotetext{a}{See Figure~\ref{field_position} for an outline.}
\tablenotetext{b}{Base position for each slit-scanning block.}
\tablenotetext{c}{Defined by [scanning direction] + [separation from the base position in arcsec].}
\tablenotetext{d}{Position angle of slit (from north to east).}
\tablenotetext{e}{Final spatial resolution on the detector, as produced by the natural seeing and optical system of CGS4.}
\tablenotetext{f}{$\hoz$ observations on 2001 August 4 (slit SE15 in field NE-1, field NE-2, and field E-0)
using the slit-jittering method described in the text).}
\end{center}
\end{table}

\end{document}